\documentclass[12pt]{iopart}     

\expandafter\let\csname equation*\endcsname\relax
\expandafter\let\csname endequation*\endcsname\relax

\usepackage[english]{babel} 
\usepackage{amsmath}
\usepackage{amssymb}
\usepackage{graphicx}
\usepackage{color}
\begin{document}

\title{ Optimal diffusive search: nonequilibrium resetting versus equilibrium
dynamics}

\author{Martin R. Evans$^{(1)}$, Satya N. Majumdar$^{(2)}$  and Kirone  Mallick$^{(3)}$}
\address{$^{(1)}$ SUPA, School of Physics and Astronomy, University of Edinburgh,
 Mayfield Road,  Edinburgh EH9 3JZ, United Kingdom \\
$^{(2)}$ Univ. Paris-Sud, CNRS, LPTMS, UMR 8626, Orsay F-01405, France \\
$^{(3)}$ Institut de Physique Th\'eorique,  CEA Saclay, Gif-Sur-Yvette F-91191,
   France}

\begin{abstract}
   We study  first-passage time problems for a diffusive particle with stochastic
 resetting with a finite rate $r$. The optimal search time is compared 
 quantitatively with that of an effective  equilibrium Langevin process
 with  the same  stationary distribution. It is shown that the intermittent,
 nonequilibrium  strategy  with  non-vanishing resetting rate is more efficient than
 the equilibrium dynamics. Our results are extended to multiparticle systems where
 a team of independent searchers, initially uniformly distributed with a given density,
 looks for a single immobile target. Both the average and the typical
 survival probability of the target are smaller  in the case of  nonequilibrium dynamics.
\end{abstract}


\maketitle

\section{Introduction}

 Stochastic search problems occur in many fields of science as well as
 in  daily life. The quest for  an optimal strategy for locating a
 target whether inanimate (such as a binding
 site  for a  protein at the molecular level \cite{Delbruck} 
 or an element in a  list) or living (such as a
 prey for a predator \cite{Bell}) has been the source of a  large
 number of different algorithms that combine observation, physical
 mechanisms and computation.  Depending on the context, search
 strategies can be very different,  leading  to a variety of
 interesting models (see the special issue  \cite{Oshanin}  devoted to
 this field of research).   One robust class of models  called {\it
 intermittent target search strategies} combine phases of slow
 motion, that allow target detection, and phases of fast motion,
 during which the searcher relocates but is not reactive (see
 \cite{BenichouRV} for a recent review). Such strategies have been
 observed at different scales: foraging animals, such as humming birds
 or bumblebees,  display intermittent search patterns \cite{Bartumeus,Stanley}.
  The  E. Coli bacteria  alternates  ballistic  moves (or `runs')
 with random changes of direction (`tumbles') in order to reach
 regions with high concentration of a chemo-attractant (chemotactic
 search) \cite{Kafri,Kafri2,Tailleur}. A  protein efficiently  localizes  a specific
 DNA sequence  by alternating  1d
 sliding phases with free 3d diffusion (`relocation
 phases'):  this   mechanism of  `facilitated diffusion', first proposed  by
 Adam and Delbr\"uck \cite{Delbruck}, enhances the association rate
 by  two orders of magnitude as compared to the diffusion limit and
 leads  to  numerical values that agree quantitatively with
 experimental results \cite{Riggs,Berg} (see  \cite{Mirny} for a
 review).

A simple model of diffusion with stochastic resetting, in which a
 Brownian particle is stochastically reset to its initial position
 with a constant rate $r$ was defined and  studied in
 \cite{EM1}. The stationary state of this process is a non-Gaussian
 distribution and  violates detailed-balance: 
 a non-vanishing steady-state  current is directed towards the  resetting
 position. This process  can be viewed as an elementary model of an
 intermittent strategy  in which the searcher, having explored its
 environment unsuccessfully for a while, returns to its initial
 position and begins a new search.  It was also shown
 in \cite{EM1} that there  exists an optimal  resetting rate
 $r^{*}$ that minimizes the average  hitting time to the
 target. Extensions to space depending rate, resetting to an random
 position with a given distribution and to a spatial distribution
 of the target were considered in~\cite{EM2}.

The effect of resetting was previously studied in
a stochastic multiplicative model of population growth where
stochastic resetting events of the population size was shown to
lead to a stationary 
power-law distributed population
size distribution~\cite{MZ}. A continuous-time random walk model
in the presence of a drift and resetting has also been studied recently~\cite{MV}.
Finally, in the context of search process, a related model has been studied
by Gelenbe~\cite{Gelenbe} where searchers are introduced stochastically
into the system: there is a single 
searcher present at a given time
with a random lifetime and when the searcher dies, a new searcher is introduced
into the system at the initial starting point.
 
In the mathematics literature, mean first-passage time for a class
of random walks with stochastic restarting events has been studied
recently from an algorithmic point of view~\cite{JP}.  

 In the present work, we focus on the simple continuous-time Brownian diffusion 
model with a nonzero resetting rate introduced in ~\cite{EM1} and 
compare quantitatively the reset dynamics
 with the equilibrium Langevin process that leads to the same
 stationary distribution. We prove  analytically that the
 non-equilibrium process is more efficient than the equilibrium one by
 showing that the optimal mean-first passage time is smaller in the
 former case. We extend our study to the multiparticle  problem by
 considering a team of searchers uniformly distributed on the line and 
 looking for an immobile target. We show that both the average
 survival probability and the typical  survival probability of the
 non-equilibrium process are smaller than the corresponding quantities
 for the  equilibrium dynamics. This shows that, at least for this
 model for which exact analytical calculations can be performed, the
 non-equilibrium strategy  systematically  defeats the  equilibrium
 behaviour.

 The outline of this work is as follows. In section~\ref{Sec:RV},
 we review the diffusion  model  with resetting and some of its basic properties,
 including  survival probability and first-passage time.
 An effective equilibrium dynamics with  the same stationary
 distribution  is defined in  section~\ref{Sec:Langevin} and it is shown
 that  the optimal mean first-passage time associated with that  dynamics
 is always larger than the  optimal time if  resetting is allowed. 
 In section~\ref{Sec:Multipart}, we investigate the survival probability of a single target
 in presence of a team of independent searchers. Average and  typical   survival probabilities
 display drastically different behaviours (algebraic vs exponential decay with time) both
 for  the nonequilibrium and  equilibrium processes. Nevertheless, it is shown 
 that  nonequilibrium dynamics  is  systematically more  efficient.
 The last section is devoted to concluding remarks.

\section{Diffusion with stochastic resetting: nonequilibrium dynamics}
 \label{Sec:RV}

  In this section, we  recall   the definition of
 the model of diffusion with stochastic resetting
  and briefly review  some basic results, derived  in \cite{EM1}.

\begin{figure}
\centerline{\includegraphics[width=10cm]{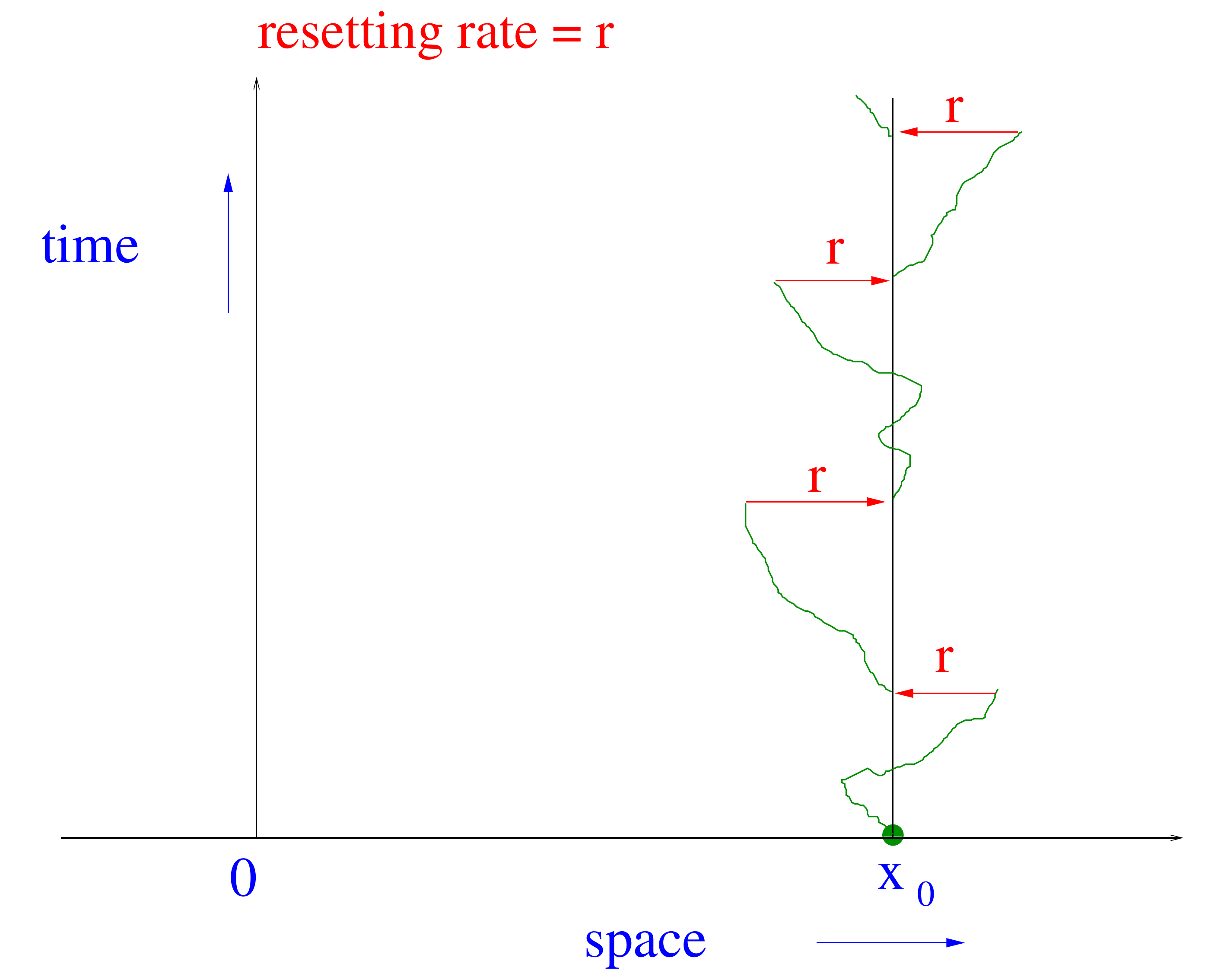}}
\caption{Schematic space-time trajectory of a one dimensional Brownian motion
that starts at $x_0$ and resets stochastically to its initial position
$x_0$ at rate $r$.}
\label{reset1.fig}
\end{figure}

 We consider a single particle on an infinite line starting at the
initial position $x_0$ at $t=0$. The position of the particle at time
$t$ is updated in a small time interval $dt$ by the following stochastic
rule~\cite{EM1}:
\begin{eqnarray}
x(t+dt) & =& x_0 \quad {\rm with\,\,prob.}\,\, r\, dt \nonumber \\
& =& x(t) + \eta(t) dt \quad {\rm with\,\,prob.}\,\, (1-r\, dt)
\label{rule.1}
\end{eqnarray}
where $\eta(t)$ is a Gaussian white noise with mean $\langle \eta(t)\rangle =0$
and the two point correlator $\langle \eta(t)\eta(t')\rangle= 2\,D\, \delta(t-t')$.
The dynamics thus  consists of a stochastic mixture of resetting to
the initial position with rate $r$ (long range move) and ordinary diffusion (short 
range move) with diffusion constant $D$ (see Fig. (\ref{reset1.fig})). Resetting 
introduces a new length
scale $\alpha_0^{-1}= \sqrt{D/r}$ in the ordinary diffusion problem.

 As shown in \cite{EM1}, the  probability density $p(x,t)$ of the particle evolves
  via the following  Fokker-Planck  equation 
\begin{equation}
\partial_t p(x,t)= D\, \partial_x^2 p(x,t)-r\, p(x,t)+r 
\delta(x-x_0) \, , 
\label{fp1}
\end{equation}
starting from the initial condition $p(x,0)=\delta(x-x_0)$. We emphasize that the dynamics
violates detailed balance manifestly: the current from a position $x$ to $x_0$
via the resetting move is not compensated by a current from $x_0$ to $x$.

The Fokker-Planck equation~(\ref{fp1}) admits a stationary solution in the $t\to 
\infty$ limit, given by 
\begin{equation}
p_{\rm st}(x) = \frac{\alpha_0}{2}\, 
\exp\left[-\alpha_0\,|x-x_0|\right]\quad {\rm
where}\,\,
\alpha_0= \sqrt{r/D}\;.
\label{st.1}
\end{equation}
Even though the stationary solution can be expressed as an effective 
Boltzmann weight: $p_{\rm st}(x)\propto \exp\left[-V_{\rm eff}(x)\right]$ with 
the effective potential $V_{\rm eff}(x)= \alpha_0\,|x-x_0|$, one should keep
in mind that this is actually a {\it nonequilibrium stationary state}, and not an
equilibrium stationary state. Indeed, in this stationary state
detailed balance is violated by  a nonzero current in the configuration 
space. Had the stationary state  been an equilibrium one, 
 the current would have  exactly vanished.

 As shown in \cite{EM1,EM2}, the  additional resetting parameter $r$ allows us to tune
the target search process to make it  more efficient.
Consider an  immobile target at the origin $x =0$ and let the searcher
undergo   diffusion with resetting dynamics specified in Eq.~(\ref{rule.1})
starting from the initial position $x_0>0$. What is the mean first-passage time 
${T}_{\rm reset}(x_0,r)$ 
to
the origin knowing that  the searcher starts  at $x_0$?
This  average time, ${T}_{\rm reset}(x_0,r)$, 
 that the searcher takes to find the target will be  taken as
a measure of the efficiency of the search process: the smaller the value of 
${T}_{\rm reset}(x_0,r)$ (for a fixed $x_0$), the better is the search.

 In order to  calculate  ${ T}_{\rm reset}(x_0,r)$, it is useful to consider the
survival probability $q_{\rm reset}(x_0,t)$ of the target given that the searcher
starts at the initial position $x_0$ and resets to $x_0$. 
 The mean-hitting time is then given quite generally by~\cite{Karlin,Redner,PaulK}, 
${T}_{\rm reset}(x_0,r)= \int_0^{\infty} t\, \left(-\partial_t 
q_{\rm reset}(x_0,t)\right)\, 
dt = \int_0^{\infty} q_{\rm reset}(x_0,t) \, dt$.

 More generally, we  define  $Q(x,x_0,t)$, the survival probability 
of the target given
that the starting position of the searcher is $x$ and its resetting position is 
$x_0$.  The initial position $x$ of the searcher  can be treated as a variable
leading to  a backward Fokker-Planck equation for $Q(x,x_0,t)$ \cite{EM1,EM2}. Eventually,
at the end of the calculation, one sets $x=x_0$ and obtains
$q(x_0,t)= Q(x=x_0,x_0,t)$. The backward 
Fokker-Planck
equation reads for $x\ge 0$
\begin{equation}
\partial_t Q(x,x_0,t)= D\, \partial_x^2 Q(x,x_0,t) 
- r\, Q(x,x_0,t)+ r\,Q(x_0,x_0,t)   \, .
\label{bfp.1}
\end{equation}
 We require the  initial condition $Q(x,x_0,t=0)=1$ for all $x>0$
 and impose 
 the following  boundary conditions: $Q(x=0,x_0,t)=0$ and $Q(x\to \infty, x_0, t)$ is 
finite. Equation~(\ref{bfp.1}) can be solved 
 explicitly in terms of the Laplace transform ${\tilde Q}(x, 
x_0,s)=\int_0^{\infty}Q(x,x_0,t)\, e^{-st}\, dt$.
In particular, setting $x=x_0$, we obtain
\begin{equation}
{\tilde q}_{\rm reset}(x_0,s)= {\tilde Q}(x=x_0,x_0,s)=
\frac{1-\exp\left(-\sqrt{\frac{r+s}{D}}\, 
x_0\right)}{s+r \, \exp\left(-\sqrt{\frac{r+s}{D}} \,
x_0\right)}  \;  , 
 \label{reset_surv.1}
\end{equation}
 thus leading to the  mean first-passage time~\cite{EM1}
\begin{equation} 
{T}_{\rm reset}(x_0,r)= \int_0^{\infty} q(x_0,t)\, dt= {\tilde q}(x_0,s=0)= 
\frac{1}{r}\left[\exp(\alpha_0\, 
x_0)-1\right],
\label{mfpt.1}
\end{equation}
where we recall that $\alpha_0=\sqrt{r/D}$.
In terms of the dimensionless parameter
$\gamma= \alpha_0 x_0=\sqrt{r/D}\, x_0$, we obtain 
\begin{equation}
{T}_{\rm reset}(x_0,\gamma)= \left[\frac{e^{\gamma}-1}{\gamma^2}\right]\, 
\frac{x_0^2}{D}\; .
\label{mfpt.2}
\end{equation}
We observe that  ${T}_{\rm reset}(x_0,\gamma)$  diverges in both the  limits
$\gamma \to 0$ (pure diffusion) and $\gamma \to \infty$  (for an infinite
resetting rate the particle is localized at $x_0$).
The  mean first-passage time has a unique minimum (see Fig. (\ref{topt.fig})) 
 at   $\gamma= \gamma_1$ where $\gamma_1$ is the solution
of $\partial_\gamma {T}=0$, i.e.,  
$\gamma_1 = 2(1-e^{-\gamma_1})$, giving $\gamma_1= 1.59362\dots$.
Hence,  for fixed values of $D$ and $x_0$, there is an optimal resetting 
rate $r=r_1= \gamma_1^2\, D/x_0^2$
that makes the search time minimum and the search process most efficient.
The  corresponding optimal mean first-passage time is given by 
\begin{equation}
{T}_{\rm reset}^{\rm opt}= \left[\frac{e^{\gamma_1}-1}{\gamma_1^2}\right]\,
\frac{x_0^2}{D}= 1.54414\dots\, \frac{x_0^2}{D} \, .
\label{opt.1}
\end{equation}

\section{An Effective Equilibrium Dynamics}
\label{Sec:Langevin}


We have seen that the resetting with diffusion leads to a 
current-carrying nonequilibrium 
stationary state   given in Eq.~(\ref{st.1}).
 This stationary distribution can  be expressed as a Boltzmann weight
$p_{\rm st}(x) =(\alpha_0/2)\, \exp\left[- V_{\rm eff}(x)\right]$ with
 an effective potential 
\begin{equation}
V_{\rm eff}(x)= \alpha_0\, |x-x_0| \, ,
 \label{def:Veff}
\end{equation}
 where  $\alpha_0$ was defined  in Eq.~(\ref{st.1}).

Let us now consider the following Langevin evolution of the particle position 
with time $t$ 
\begin{equation}
\frac{dx}{dt}= - B\, \partial_x V_{\rm eff}(x) + \eta(t)
\label{lange.1}
\end{equation}
where $B$ is the amplitude of the external force and $\eta(t)$ is
the same Gaussian white noise as in the previous case, i.e., with
$\langle \eta(t)\rangle =0$ and $\langle \eta(t)\eta(t')\rangle = 2\,D\, 
\delta(t-t')$. Note that we have chosen the same diffusion constant
$D$ as in the resetting case  to reflect the fact that without
the reset (in the former case) or without the external potential (in
the Langevin case), this equation  describes ordinary diffusion
with the same diffusion constant $D$ in both cases. The  corresponding Fokker-Planck equation
for the probability density $P(x,t)$ of the particle is 
\begin{equation}
\partial_t P(x,t)= D\, \partial_x^2 P(x,t) + \partial_x\left[B\, \left(\partial_x 
V_{\rm 
eff}(x)\right)\, P(x,t) \right]\,.
\label{fp.2}
\end{equation}
Eq. (\ref{fp.2}) can be expressed as a continuity equation, $\partial_t 
P=-\partial_x J$ where the current density $J(x,t)= -D \partial_x P -B\, (\partial_x 
V_{\rm eff}(x))\, P$. The system will then reach a stationary state at long times
and if we set the current in the stationary state to be $0$, we arrive at
the equilibrium Gibbs-Boltzmann solution which reads, $P_{\rm eq}(x)={\cal N}\, 
\exp\left[- (B/D)\, V_{\rm eff}(x)\right]$, where $\cal N$ is the normalization 
constant. By choosing $B=D$
and $V_{\rm eff}(x)=\alpha_0\, |x-x_0|$,  we can  engineer  a zero current equilibrium state which has the same weight as the 
current carrying nonequilibrium
stationary state in the resetting case, i.e.,
\begin{equation}
P_{\rm eq}(x)= p_{\rm st}(x) = \frac{\alpha_0}{2}\, \exp\left[-\alpha_0\, 
|x-x_0|\right]\quad 
{\rm where}\,\, \alpha_0= \sqrt{r/D}\,.
\label{eq_state}
\end{equation}


The following natural question then arises. Consider the target
search problem, where we have an immobile target at the origin.
In the previous section, the  searcher was performing  normal diffusion with
stochastic resetting to its initial position $x_0$.  Now, 
suppose  that  the searcher undergoes instead  the Langevin dynamics
as in Eq. (\ref{lange.1}) with the choice $B=D$ and $V_{\rm eff}(x)=\alpha_0\, 
|x-x_0|$ that guarantees that both dynamics lead to the same steady state.
One is then tempted to 
compare the efficiency of the nonequilibrium search process with diffusion and 
reset to the Langevin search process where the searcher's position evolves
via Eq. (\ref{lange.1}). Which process  is the  more efficient?

To address this question, we need to compute the mean first-passage time
${T}_{\rm lange}(x_0,r)$ to the origin of the Langevin process
in Eq. (\ref{lange.1}). With the choice $B=D$ and $V_{\rm eff}(x)=\alpha_0\, 
|x-x_0|$
where $\alpha_0=\sqrt{r/D}$, the Langevin equation reads
\begin{equation}
\frac{dx}{dt}= -\alpha_0\, D\, {\rm sgn}(x-x_0) +\eta(t)
\label{lange.2}
\end{equation}
where ${\rm sgn}(z)$ denotes the sign of $z$.
We  define $Q_{\rm lange}(x,x_0,t)$,  the probability
that the searcher, starting at the initial position $x\ge 0$ and evolving via
Eq. (\ref{lange.2}), does not reach the origin (the target) up to time $t$.
Equivalently,  $Q_{\rm lange}(x,x_0,t)$ is the survival probability of the target
up to time $t$ under the Langevin dynamics of the searcher. 
Treating the initial position $x$ as a variable,  $Q_{\rm lange}(x,x_0,t)$ satisfies
a  backward Fokker-Planck equation  \cite{Karlin,Redner}
\begin{equation}
\partial_t Q_{\rm lange}(x,x_0,t)= D\, \partial_x^2 Q_{\rm lange}(x,x_0,t)- 
\alpha_0\, D\, {\rm 
sgn}(x-x_0)\,\partial_x Q_{\rm lange}(x,x_0,t)
\label{bfp.2}
\end{equation}
which holds for all $x\ge 0$ with the initial condition
$Q_{\rm lange}(x,x_0,t=0)=1$ for all $x>0$ and 
the boundary conditions: (i) $Q_{\rm 
lange}(x=0,x_0,t)=0$ for all $t$ (absorbing boundary at the origin)
and (ii) $Q_{\rm lange}(x\to \infty, x_0,t)=1$.

To solve Eq. (\ref{bfp.2}), we  consider the Laplace transform
${\tilde Q}_{\rm lange}(x,x_0,s)= \int_0^{\infty} Q_{\rm lange}(x,x_0,t)\, e^{-st}\, 
dt$, which,  taking into account  the initial condition, satisfies
\begin{equation}
-1+ s\, {\tilde Q}_{\rm lange}(x,x_0,s)= D\, \frac{d^2 {\tilde Q}_{\rm 
lange}}{dx^2}- 
\alpha_0\, D\, {\rm sgn}(x-x_0)\, \frac{d {\tilde Q}_{\rm lange}}{dx}\,.
\label{lap.1}
\end{equation}
Making a shift ${\tilde Q}_{\rm lange}(x,x_0,s)= 1/s + {\tilde F}(x,x_0,s)$,
one finds a homogeneous differential equation for ${\tilde F}$ in $x\ge 0$
\begin{equation}
D\, \frac{d^2 {\tilde F}}{dx^2}- \alpha_0\, D\, {\rm sgn}(x-x_0)\, \frac{d {\tilde 
F}}{dx} - s\, {\tilde F}=0 
\label{lap.2}
\end{equation}     
with the boundary conditions: (i) ${\tilde F}(x=0,x_0,s)= -1/s$ and (ii) ${\tilde 
F}(x\to \infty, x_0,s)=0$.  

Taking  the boundary condition (ii)  into account, we obtain
\begin{eqnarray}
{\tilde F}(x,x_0,s)&=& A_1 \, 
\exp[-\mu_2\, (x-x_0)]\quad {\rm for}\, x>x_0 \label{xgt} \\
{\tilde F}(x,x_0,s)&=& 
B_1\, \exp\left[-\mu_1 (x-x_0)\right] + B_2\, \exp\left[\mu_2 
(x-x_0)\right]\quad {\rm for}\, x<x_0 \, ,
\label{xlt} 
\end{eqnarray}
where 
\begin{equation}
\mu_1= (\sqrt{\alpha_0^2+4s/D}+\alpha_0)/2\, , \quad {\rm and  } \quad 
\mu_2= (\sqrt{\alpha_0^2+4s/D}-\alpha_0)/2 \, .
\label{def:mu}
\end{equation}    
From the boundary condition (i) at $x=0$ and using
that  ${\tilde F}(x,x_0,s)$ and its first derivative
$\partial_x {\tilde F}(x,x_0,s)$ are continuous at $x=x_0$, 
the three unknown constants $A_1$,  $B_1$ and $B_2$ are determined
\begin{equation} 
B_1= \frac{2\mu_2}{\mu_1+\mu_2}\, A_1; \quad B_2= 
\frac{\mu_1-\mu_2}{\mu_1+\mu_2}\, 
A_1; \quad {\rm and}\,\, A_1= -\frac{1}{s}\, 
\frac{(\mu_1+\mu_2)}{[2\mu_2e^{\mu_1\,x_0}+ (\mu_1-\mu_2)\, e^{-\mu_2\, x_0}]}\,.
\label{constants}
\end{equation}
Finally, inserting  $x=x_0$ leads to 
\begin{equation}
{\tilde q}_{\rm lange}(x_0,s)= {\tilde Q}_{\rm lange}(x=x_0,x_0,s)= 
\frac{1}{s}\,\left[1- \frac{(\mu_1+\mu_2)}{\left[2\mu_2e^{\mu_1\,x_0}+ 
(\mu_1-\mu_2)\, 
e^{-\mu_2\, x_0}\right]}\right] \, . 
\label{lange_surv.1}
\end{equation}
The mean first-passage time is then given by
$${T}_{\rm lange}(x_0,r)= \int_0^{\infty} t\, (-\partial_t 
q_{\rm lange}(x_0,t))\,dt=\int_0^{\infty} q_{\rm lange}(x_0,t)\, dt= {\tilde q}_{\rm 
lange}(x_0,s=0) \, . $$
Taking the $s\to 0$ limit in Eq.~(\ref{lange_surv.1}) leads us to 
\begin{equation}
{T}_{\rm lange}(x_0,r)= \frac{1}{\alpha_0^2\, D}\,\left[2\,(e^{\alpha_0\, 
x_0}-1)-\alpha_0\,x_0\right]= \left[\frac{2(e^{\gamma}-1)-\gamma}{\gamma^2}\right]\, 
\frac{x_0^2}{D}
\label{mfpt_lange.1}
\end{equation}
where $\gamma= \alpha_0 x_0=\sqrt{r/D}\, x_0$ is the same dimensionless parameter
as defined above (see Eq.~(\ref{mfpt.2})). 
We can now  compare the result in Eq. (\ref{mfpt_lange.1}) with that of 
 the resetting case,  Eq. (\ref{mfpt.2}).  Using the fact that 
 $e^{\gamma}-1 \ge \gamma$, we see that 
 ${T}_{\rm lange}(x_0,r) \ge  {T}_{\rm reset}(x_0,r)$ for fixed values
 of $x_0$ and $D$.

\begin{figure}
\centerline{\includegraphics[width=10cm]{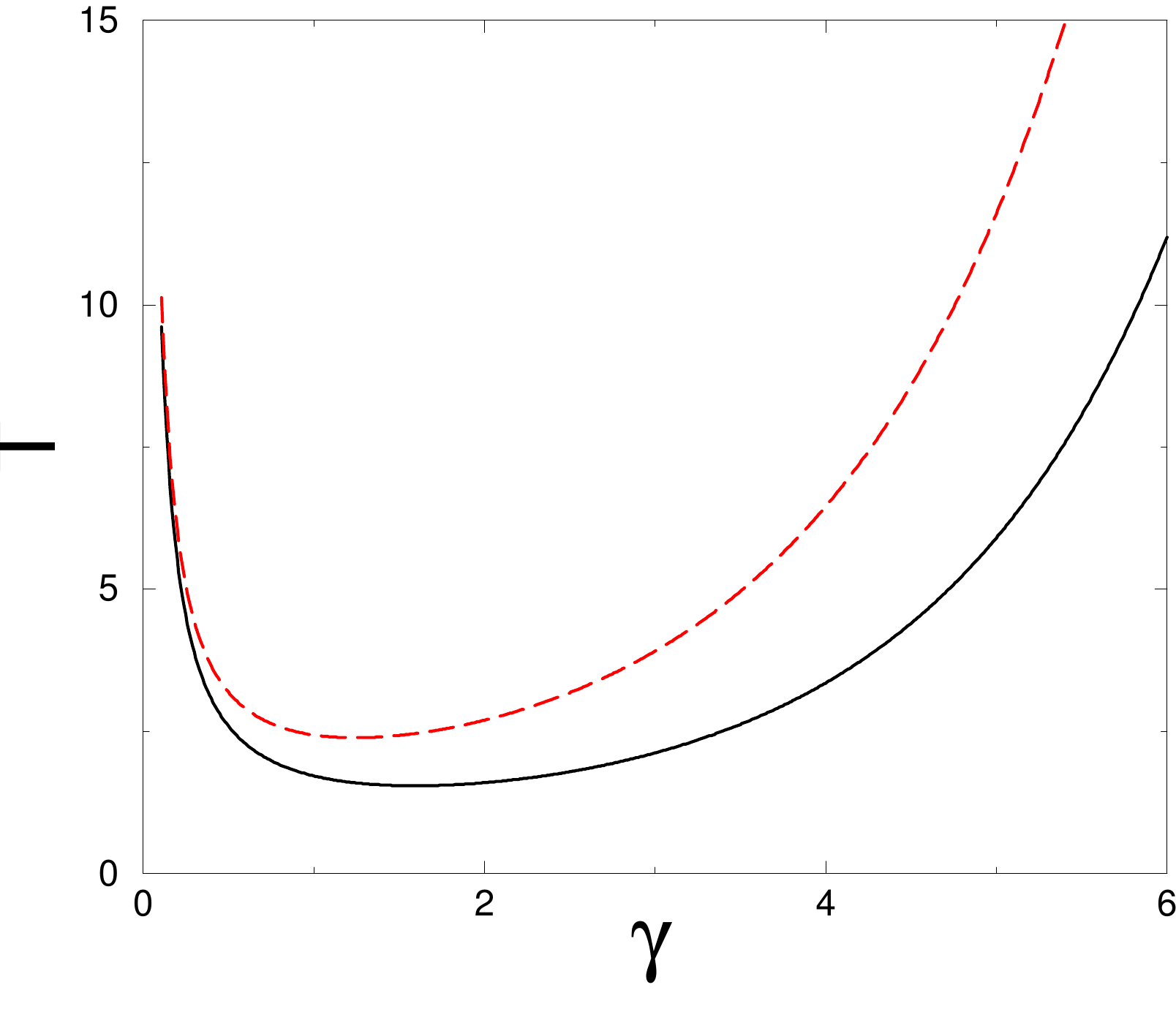}}
\caption{Mean first-passage time $T$ plotted versus the dimensionless
parameter $\gamma$ (setting $x_0=1$ and $D=1$) for the
nonequilibrium reset dynamics (shown by the solid (black) line) (see Eq. 
(\ref{mfpt.2}))
and the equilibrium Langevin dynamics (dashed (red) line) (see Eq. 
(\ref{mfpt_lange.1})). Evidently, the optimal (minimum) $T$ 
is higher in the equilibrium Langevin case.} 
\label{topt.fig}
\end{figure}

 The minimum value of 
${T}_{\rm lange}(x_0,r)$ is obtained (see Fig. (\ref{topt.fig})) for 
$\gamma=\gamma_2$ where
$\gamma_2$ is given by the solution of $\partial_\gamma {T}_{\rm 
lange}(x_0,r)=0$, i.e., it is the positive root of the equation
$2(\gamma_2-2)\, e^{\gamma_2} +\gamma_2+4=0$, leading to 
$\gamma_2= 1.24468\ldots$. Thus the optimal mean first-passage time with
Langevin dynamics of the searcher is given by
\begin{equation}
{T}_{\rm lange}^{\rm opt}= 
\left[\frac{2(e^{\gamma_2}-1)-\gamma_2}{\gamma_2^2}\right]\,
\frac{x_0^2}{D}= 2.38762\dots \, \frac{x_0^2}{D}\,.
\label{opt.2}
\end{equation}
Comparing with  the corresponding result in Eq. (\ref{opt.1}) for the reset
dynamics, we conclude that 
\begin{equation}
\frac{{T}_{\rm reset}^{\rm opt}}{{T}_{\rm lange}^{\rm opt}}= 
\frac{1.54414\dots}{2.38762\dots}= 0.646728\dots \le 1  \, .
\label{ratio.1}
\end{equation}
 This shows that  the search process via the nonequilibrium  diffusion combined with reset
mechanism is significantly  more efficient than the equilibrium Langevin dynamics
of the searcher although the stationary distributions induced by both dynamics are the same: by this measure
the  nonequilibrium strategy  beats the  equilibrium dynamics. 




\section{Multiparticle problem: Nonequilibrium vs. Langevin dynamics}
\label{Sec:Multipart}

In the multiparticle version of the search process, we have a single immobile target
at the origin and a team of searchers which are initially uniformly distributed
on the line with uniform density $\rho$ (see \cite{SidPaul} for a colorful example).
  The searchers are independent of 
each other and the position of each searcher evolves stochastically (identical 
in law for all searchers)
starting at its own initial position. This stochastic process, for the moment,
is general. When any of the searchers finds the target, the search process is terminated.
Let $P_s(t)$ denote the survival probability of the target up to time $t$, i.e,
the probability that the target has not been found up to $t$ by any of the searchers.
To set the problem, we consider $N$ searchers initially distributed
uniformly in a box $[-L/2,L/2]$ of size $L$ and 
 will  eventually take the limit $N\to \infty$, $L\to \infty$, but keeping the
density $\rho=N/L$ fixed. Let $x_i$ denote the initial position
of the $i$-th searcher. Thus, $x_i$ is a random variable uniformly distributed
in the box $[-L/2,L/2]$. 
Since the searchers are independent, it follows  that
for a given set of initial positions of the searchers $\{x_i\}$, 
\begin{equation}
P_s(t)= \prod_{i=1}^N q(x_i,t)
\label{mp.1}
\end{equation}
where $q(x_i,t)$ is the survival probability of the target due to a single
searcher starting at $x_i$. For both (i) diffusion with resetting dynamics
and (ii) Langevin dynamics in an effective potential $V_{\rm eff}(x)=\alpha_0\,|x-x_0|$,
the Laplace transforms of $q(x_i,t)$ are given respectively in Eqs. (\ref{reset_surv.1})
and (\ref{lange_surv.1}). The explicit dependence of $P_s(t)$ on $x_i$'s has been
suppressed in Eq. (\ref{mp.1}) for notational convenience.

We first compute the average survival probability (averaged over the initial
positions of the searchers) following Ref.~\cite{EM1}. Taking the average of  Eq. (\ref{mp.1})
we obtain
\begin{eqnarray}
\langle P_s(t)\rangle &= & \prod_{i=1}^N [1- \langle (1-q(x_i,t)\rangle] \nonumber \\
&= & \prod_{i=1}^N \left[1- \frac{1}{L}\int_{-L/2}^{L/2} 
\left(1-q(x_i,t)\right)\,dx_i\right] \nonumber \\
& \to & \exp\left[- \rho \int_{-\infty}^{\infty} [1-q(x,t)]\, dx\right]
\label{av_surv.1}
\end{eqnarray}
where in the last step we have exponentiated for large $L$ and then taken the
thermodynamic limit $N\to \infty$, $L\to \infty$ but keeping the ratio
$\rho=N/L$ fixed. Using further the symmetry $q(x,t)= q(-x,t)$, one finally obtains
the general expression of the average survival probability in the multiparticle
system
\begin{equation}
\langle P_s(t)\rangle= \exp\left[-2\, \rho\, \int_0^{\infty} [1-q(x,t)]\, dx\right]\, .
\label{av_surv.2}
\end{equation}
Actually, this result is rather general (see e.g. \cite{Tachiya}) and holds for any stochastic process, the 
only assumption being that all searchers undergo the same stochastic process.
If one can estimate the survival probability $q(x,t)$ of the target at the
origin for a single searcher starting at $x$, one has an exact formula
for the average survival probability in the multiparticle problem.
The quantity $M(t)=\int_0^{\infty} [1-q(x,t)]\, dx$ has the interpretation 
as the expected maximum up to time $t$ of the single particle process
(starting at the origin) and the general formula in Eq. (\ref{av_surv.2})
and its discrete-time analogue
have been used recently to compute exactly the average survival
probability for a number of processes including continuous-time random walks
as well as discrete-time L\'evy flights~\cite{FM}.

Similarly, the typical survival probability can be estimated~\cite{EM1}
 by first taking logarithm
of Eq. (\ref{mp.1}), followed by averaging over the initial positions
and finally reexponentiating
\begin{eqnarray}
P_s^{\rm typ}(t) &= &\exp\left[ \sum_{i=1}^N \langle \ln [q(x_i,t)] \rangle\right] 
\nonumber \\
&\to & \exp\left[ 2\, \rho\, \int_0^{\infty} \ln [q(x,t)]\, dx\right]\, .
\label{typ_surv.2}
\end{eqnarray}
Thus, if we know the survival probability $q(x,t)$ in the single searcher case,
we can use the two exact formulae in Eqs. (\ref{av_surv.2}) and (\ref{typ_surv.2})
to estimate respectively the average and the typical survival probability 
of the target in the multiparticle case.

\subsection{Average survival probability: nonequilibrium vs. equilibrium}

Here,  we  analyze the asymptotic large $t$ behavior of $\langle P_s(t)\rangle$
in Eq. (\ref{av_surv.2}) and compare the expression obtained for the resetting dynamics
with the one for the Langevin dynamics.

 Let us denote
\begin{equation}
M(t) =\int_0^{\infty} [1-q(x,t)]\, dx\quad {\rm so}\,\, {\rm that}\,\, \langle 
P_s(t)\rangle = \exp\left[-2\,\rho\, M(t)\right]
\label{def_mt}
\end{equation}
In terms of the  Laplace transform  of $M(t)$,  
${\tilde M}(s)= \int_0^{\infty} M(t)\, e^{-s\, t}\, dt$,   Eq. (\ref{def_mt}) reads
\begin{equation}
{\tilde M}(s)= \int_0^{\infty} \left[\frac{1}{s}- {\tilde q}(x,s)\right]\, dx
\label{ms_def}
\end{equation}
We now  consider the two cases (i) diffusion with reset and (ii) Langevin dynamics
separately.

\vskip 0.3cm
 \subsubsection{Nonequilibrium case (diffusion with reset)}
Substituting ${\tilde q}_{\rm reset}(x,s)$ from 
Eq. (\ref{reset_surv.1}) in Eq. (\ref{ms_def})
and performing the integration over $x$,  one obtains exactly~\cite{EM1}
\begin{equation}
{\tilde M}(s)=  \frac{\sqrt{D(r+s)}}{r\,s}\, \ln\left(\frac{r+s}{s}\right)\,.
\label{ms_reset.1}
\end{equation}
The  large $t$ behavior of $M(t)$ will be derived from 
small $s$ behavior of ${\tilde M}(s)$: 
\begin{equation}
 \hbox{ As }  \quad s\to 0\,,  \quad  \quad  
{\tilde M}(s)\approx \sqrt{\frac{D}{r}}\,\left[-\frac{\ln s}{s}  
+\frac{\ln (r)}{s}+ \dots\right]\, .
\label{ms_reset.2}
\end{equation}  
Inverting
the Laplace transform, we  find that the leading asymptotic 
behavior of $M(t)$ for large $t$  is given by 
\begin{equation}
M(t) \approx \sqrt{\frac{D}{r}}\,\ln t + \sqrt{\frac{D}{r}}\, \ln(r) +\dots
\label{mt_reset.1}
\end{equation}
Thus,  the average survival probability from Eq. (\ref{def_mt}) 
decays algebraically for large $t$~\cite{EM1}
\begin{equation}
\langle P_s^{\rm reset} (t) \rangle \approx a_1\, t^{-\theta_1}\quad {\rm where}\,\, 
\theta_1= 2\,\rho\, \sqrt{D/r}\quad {\rm and}\,\, a_1=\exp\left[-\theta_1\, \ln 
r\right]=r^{-\theta_1}\, .
\label{av_surv_reset.1}
\end{equation}   
   
\vskip 0.3cm
\subsubsection{Equilibrium case (Langevin dynamics)}
Here, we use the expression~(\ref{lange_surv.1})
 of  ${\tilde q}_{\rm lange}(x,s)$  in Eq.~(\ref{ms_def}), 
  perform the integration over $x$ exactly and get 
\begin{equation}
{\tilde M}(s)=  \left[\frac{\mu_1+\mu_2}{2\mu_1\mu_2\, s}\right]\,\, 
_2F_1\left(1,\frac{\mu_1}{\mu_1+\mu_2}, 
\frac{2\mu_1+\mu_2}{\mu_1+\mu_2},\frac{\mu_2-\mu_1}{2\mu_2}\right)
\label{ms_lange.1} 
\end{equation}
where $\mu_{1,2}$  are defined in Eq.~(\ref{def:mu}), $\alpha_0$ in Eq.~(\ref{st.1})
and  the hypergeometric function  $_2F_1(a,b,c,z)$ is  given by 
\begin{equation}
_2F_1(a,b,c,z)= 1+ \frac{ab}{c}\,z+ \frac{a(a+1)b(b+1)}{c(c+1)}\, \frac{z^2}{2!}
+ \frac{a(a+1)(a+2)b(b+1)(b+2)}{c(c+1)(c+2)}\, \frac{z^3}{3!}+\ldots
\label{hyper_def}
\end{equation}
Expanding $\mu_{1,2}$ for small $s$, keeping terms up to $O(s)$ and using $\alpha_0^2\, 
D=r$,  one gets
\begin{equation}
{\tilde M}(s)\approx \frac{\alpha_0\, D}{2\,s^2}\,\left[1+ \frac{2s}{r}+\cdots\right]\, 
_2F_1\left[1, 1- \frac{s}{r}, 2-\frac{s}{r},- 
\frac{r}{2s}-\frac{1}{2}+\frac{s}{2r}\right]
\label{ms_lange.2}
\end{equation}
To make further progress, we use the following identity
$_2F_1(1,1,2,-z)=\ln(1+z)/z $ which follows from the definition
Eq. (\ref{hyper_def}). Expanding for small $s$, we find the following leading 
order behavior of
${\tilde M}(s)$ from Eq. (\ref{ms_lange.2})
\begin{equation}
{\tilde M}(s)= -\frac{1}{\alpha_0 s}\, \ln (s) + \frac{1}{\alpha_0\, s}\,\ln 
(r/2)+ 
\ldots
\label{ms_lange.3}
\end{equation}
where $\ldots$ correspond to lower order terms that vanish as $s\to 0$.
This indicates that for large $t$, 
\begin{equation}
M(t) \approx \frac{1}{\alpha_0}\, \ln(t) + \frac{1}{\alpha_0}\,\ln (r/2)+ \ldots
\label{Mtlarge}
\end{equation}
Hence, from 
Eq. (\ref{def_mt}), we obtain  the average survival probability
\begin{equation}
\langle P_s^{\rm lange} (t) \rangle \approx a_2\, t^{-\theta_2}\quad {\rm where}\,\,
\theta_2= \theta_1=2\,\rho\, \sqrt{D/r} \quad {\rm and}\,\, a_2=e^{-\theta_1\, 
\ln(r/2)}= (r/2)^{-\theta_1}
\label{av_surv_lange.1}
\end{equation}
Thus the power law exponent $\theta_2$ characterizing the algebraic decay 
of the average survival probability in the Langevin 
case is identical to the nonequilibrium case, though the amplitude 
$a_2=2^{\theta_1}\, a_1$ is larger than $a_1$.
Hence, for large $t$, the average survival probability in the Langevin case
is greater than that of the nonequilibrium case
\begin{equation}
\langle P_s^{\rm reset} (t)\rangle \le \langle  P_s^{\rm lange} (t)\rangle \, .
\label{ineq.1}
\end{equation}
Thus, we conclude that on average the target is found faster in the
nonequilibrium case than in the equilibrium one.

\subsection{Typical survival probability: nonequilibrium vs. equilibrium}

We  now analyze the asymptotic large $t$ behavior of the typical survival probability, 
 defined in Eq. (\ref{typ_surv.2}). We define
\begin{equation}
W(t) =\int_0^{\infty} \ln [q(x,t)]\, dx\quad {\rm so}\,\, {\rm that}\,\, 
P_s^{\rm typ}(t) = \exp\left[2\,\rho\, W(t)\right] \, .
\label{def_wt}
\end{equation}
We recall  that the Laplace transform of $q(x,t)$ is denoted by
${\tilde q}(x,s)$ and its expressions are given in Eqs. (\ref{reset_surv.1}) 
and (\ref{lange_surv.1}) respectively for (i) the diffusion with reset dynamics
and (ii) the equilibrium Langevin dynamics in the effective potential
$V_{\rm eff}(x)=\alpha_0\, |x-x_0|$. The two cases will be considered separately.

\vskip 0.3cm  
 \subsubsection{Nonequilibrium case (diffusion with reset)}
 To  analyze the asymptotic large $t$ behavior of $W(t)$, we need to know
 how $q(x,t)$ behaves for large $t$. The Laplace transform
of $q(x,t)$,  given in   Eq.~(\ref{reset_surv.1}), 
 has a
pole at $s=s_1$ (for fixed $r$ and fixed $x$) which satisfies 
\begin{equation}
s_1+ r \, \exp\left[-\sqrt{\frac{r+s_1}{D}}\, x\right]=0 \, .
\label{reset_root.1}
\end{equation}
Clearly $s_1=s_1(x)$ depends implicitly on $x$ and 
one must have  $s_1(x)<0$. Then, to leading order for large $t$,
it follows from  Laplace inversion  that
\begin{equation}
q(x,t) \sim \exp \left[ s_1(x)\, t\right]= \exp \left[- |s_1(x)|\, t \right] \, .
\label{qasymp_reset.1}
\end{equation}
Consequently, from Eq. (\ref{def_wt}), we find  to leading order for large $t$
\begin{equation}
W(t) \approx - \left[\int_0^{\infty} |s_1(x)|\, dx\right] \, t \; .
\label{wtasymp_reset.1}
\end{equation}
Hence,  the typical survival probability decays exponentially for large $t$
as
\begin{equation}
P_s^{\rm typ}(t) \sim \exp\left[-2\,\rho\,\kappa_1\, t\right] \quad 
{\rm 
where}\,\, 
\kappa_1 = \int_0^{\infty}|s_1(x)|\, dx \, .
\label{typ_surv_reset.1}
\end{equation}

To compute $s_1(x)$, it is useful to first define $s_1(x)= -r (1-u)$, so
that Eq. (\ref{reset_root.1}) reads, in terms of $u$ and the
dimensionless length $z=\alpha_0\, x$
\begin{equation}
u-1 +\exp\left[-\sqrt{u}\, z\right]=0\, .
\label{u1_def}
\end{equation}
Clearly, as $z\to \infty$, $u(z)\to 1$ and as $z\to 0$, $u(z)\to 0$. Hence
\begin{equation}
\kappa_1= \int_0^{\infty} |s_1(x)|\, dx= \frac{r}{\alpha_0}\, \int_0^{\infty} 
[1-u(z)]\, dz\, .
\label{kappa_reset.1}
\end{equation}
The idea then is to transform the integral over $z$ to an integral over
$u$. We then use $dz= du/|u'(z)|$ where $u'(z)= du(z)/dz$.
The derivative can be easily computed from Eq. (\ref{u1_def}).
Expressing Eq. (\ref{u1_def}) as $z= - \ln(1-u)/\sqrt{u}$ 
and taking derivative, we get
\begin{equation}
\frac{dz}{du}= \frac{1}{2u^{3/2}}\, \ln(1-u) + \frac{1}{\sqrt{u}\,(1-u)}\, .
\label{kappa_reset.2}
\end{equation}
Hence this gives, using $\alpha_0=\sqrt{r/D}$,
\begin{equation}
\kappa_1= \sqrt{r\,D}\, \int_0^{\infty} [1-u(z)]\, dz\,= \sqrt{r\,D}\, \int_0^1 
(1-u)\left[\frac{1}{2u^{3/2}}\, \ln(1-u) + \frac{1}{\sqrt{u}\,(1-u)}\right]\, 
du\, .
\label{kappa_reset.3}
\end{equation}
The integral in Eq. (\ref{kappa_reset.3}) can be done explicitly to give~\cite{EM1}
\begin{equation}
\kappa_1= 4\, (1-\ln 2)\, \sqrt{r\,D}\,.
\label{kappa_reset.4}
\end{equation}
Hence finally we find
\begin{equation}
P_s^{\rm typ}(t) \sim \exp\left[-8\,(1-\ln 2)\,\rho\,\sqrt{rD}\, t\right] 
= \exp\left[-(2.45482\dots)\, \rho\,\sqrt{rD}\, t\right]\,.
\label{typ_surv_reset.2}
\end{equation}
   As explained in ~\cite{EM1}, the fact that the average and the typical survival probabilities
  have different behaviours in the large time limit is a consequence of the memory
 of the initial conditions in the diffusion process with resetting.
\vskip 0.3cm  
 \subsubsection{Equilibrium case (Langevin dynamics)}
In this case, we proceed in exactly the same  way as  the nonequilibrium case, except
that to evaluate $W(t)= \int_0^{\infty} \ln [q(x,t)]\, dx$ in Eq. (\ref{def_wt}), we need
 the expression~(\ref{lange_surv.1}) of the 
 Laplace transform of $q_{\rm lange}(x,t)$. The function ${\tilde q}_{\rm lange}(x,s)$
has a pole at $s=s_2(x)$ which is  a
root of the equation (with fixed $r$ and fixed $x$)
\begin{equation}
2\, \mu_2(s_2)\, \exp[\mu_1(s_2)\, x]+ (\mu_1(s_2)-\mu_2(s_2))\, \exp[-\mu_2(s_2)\, x]=0 \, , 
\label{lange_root.1}
\end{equation}
where we emphasize  that $\mu_{1,2}(s)$  defined in Eq.~(\ref{def:mu})  depend
on $s$.  Because  $\mu_{1} >  \mu_{2}$, one must have $\mu_{2}(s_2)<0$
 and therefore  $s_2(x)<0$ to fulfil this equation.
 Thus, we find  $q(x,t)\sim \exp[-|s_2(x)|\, t]$ for large 
$t$ and  $W(t)\sim -\left[\int_0^{\infty}|s_2(x)|\, dx\right]\, t$. 
Consequently
\begin{equation}
P_s^{\rm typ}(t)= \exp\left[2\, \rho\, W(t)\right]\sim \exp\left[- 2\,\rho\, \kappa_2\, 
t\right]
\label{typ_surv_lange.1}
\end{equation}    
where
\begin{equation}
\kappa_2= \int_0^{\infty}|s_2(x)|\, dx\, .
\label{kappa2.1}
\end{equation}
To compute $\kappa_2$, we reorganize Eq. (\ref{lange_root.1}) slightly (using
the explicit expressions of $\mu_1$ and $\mu_2$)
and express it in terms of the dimensionless length $z=\alpha_0\, x$
as
\begin{equation}
\sqrt{1+ 4s_2/r}-1+\exp\left[-\sqrt{1+4s_2/r}\, z\right]=0\, .
\label{lange_root.2}
\end{equation}
where we have used $\alpha_0^2 D=r$. Let us further define $v= 1+ 4s_2/r$ in terms of 
which
Eq. (\ref{lange_root.2}) reads
\begin{equation}
\sqrt{v}-1+\exp\left[-\sqrt{v}\, z\right]=0\, .
\label{lange_root.3}
\end{equation}
Hence, from Eq. (\ref{kappa2.1}) we get
\begin{equation}
\kappa_2= \frac{\sqrt{r\,D}}{4}\, \int_0^{\infty} [1-v(z)]\, dz
\label{kappa2.2}
\end{equation}
where $v(z)$ is the solution of Eq. (\ref{lange_root.3}). 
Now  the solution of Eq. (\ref{lange_root.3})
has two branches: $v=0$, $0 \leq z < 1$ and a branch with $v(z)>0$ for $z\geq 1$.
Thus (\ref{kappa2.2}) becomes
\begin{equation}
\kappa_2= \frac{\sqrt{r\,D}}{4}\left[ 1
+ \int_1^{\infty} [1-v(z)]\, dz\right]
\label{kappa2.3}
\end{equation}
To compute the integral in
(\ref{kappa2.3}), we use the same 
trick as above. 
\begin{equation}
\int_1^{\infty} [1-v(z)]\, dz
= \int_0^{1} [1-v]\,\frac{d z(v)}{dv}\, dv\;.
\label{int}
\end{equation}
where $z(v)=- \ln(1-\sqrt{v})/\sqrt{v}$,
which follows from  (\ref{lange_root.3}).
Here, the integral
(\ref{int})
is easily evaluated by integration by parts
\begin{eqnarray}
\int_0^{1} (1-v)\,\frac{dz(v)}{dv}\, dv
&=& \left[ (1-v)z(v) \right]_{v=0}^{1} + \int_0^{1} z(v)\, dv \nonumber \\
& =& -1  - \int_0^1\frac{\ln(1- v^{1/2})}{v^{1/2}}\, dv
= 1\;.
\end{eqnarray}
Therefore from (\ref{kappa2.3}) we have
\begin{equation}
\kappa_2= \frac{\sqrt{r\,D}}{2}\, .
\label{kappa2.4}
\end{equation} 
Hence finally we obtain
\begin{equation}
P_s^{\rm typ}(t) \sim \exp\left[-\rho\,\sqrt{rD}\, t\right]
\,.  
\label{typ_surv_lange.2}
\end{equation} 

Comparing this result to Eq. (\ref{typ_surv_reset.2}), we  find
that for the multiparticle case as well, the typical survival probability of the target
in the equilibrium Langevin case is larger than the nonequilibrium dynamics
of diffusion with resetting
\begin{equation}
P_s^{\rm typ}(t)\big|_{\rm reset}\le P_s^{\rm typ}(t)\big|_{\rm lange}\, .
\label{compare_typ}
\end{equation}
This means that the target is found faster in the nonequilibrium case than
the equilibrium Langevin case. In other words, the 
target search process by multiple searchers, as in the case of a single searcher,
is more efficient when the searcher undergoes nonequilibrium diffusion and reset 
dynamics, rather  than the equilibrium  Langevin dynamics.

\section{Concluding Remarks}

   Equilibrium thermodynamics teaches us that dissipation is reduced
 when  a system  remains close to an 
 equilibrium state and that the transformations that affect  it 
 are  quasistatic and reversible. Such a statement of local optimum can not be
 taken as a paradigm: it can be advantageous in some circumstances 
 to be driven  away from  equilibrium,  by creating
 non-vanishing stationary currents that  break detailed balance and 
 time-reversal invariance.
 
   Optimal search problems provide us with concrete examples in which nonequilibrium
 can defeat equilibrium. In the present work, we have undertaken a systematic
 comparison between two related search strategies. We consider a  predator who performs
 an intermittent search by alternating phases of free diffusion and  resetting
 jumps. The invariant  distribution corresponding to such a dynamics is
  a nonequilibrium stationary state in which
 a non-vanishing probability current is  directed towards the  resetting position.
 However, one can readily  define a fictitious equilibrium Langevin dynamics leading 
 to the same stationary distribution (but with different local transition rates,
 leading to a vanishing current in the steady state). For the resetting model,
 this amounts to defining an effective attractive potential centered on the 
 resetting position.  The hitting-times to an immobile target can be exactly
 calculated in both dynamics (nonequilibrium vs equilibrium) and compared. 
 The same problem can be addressed in the case of a team of independent searchers;
 here the typical and the average survival probabilities differ.  Nevertheless,
 in all the cases we have studied in one dimension, we have shown that the 
nonequilibrium strategy
 is advantageous: the intermittent search is always more efficient than  equilibrium
 diffusion. It would be interesting to see if the same conclusion holds
in higher dimensions as well and in more realistic search scenarios
where, for example, there is a time penalty for resetting.


\vspace{1cm}

\begin{thebibliography}{99}


 \bibitem{Delbruck} G. Adam and M. Delbr\"uck, { Reduction of dimensionality
 in biological diffusion processes}, in {\it Structural Chemistry and Molecular 
   Biology,  A. Rich and N. Davidson Eds.}
(W.H. Freeman and Company, San Francisco; London, 1968).

\bibitem{Bell} W. J. Bell, {\it Searching behaviour: the behavioural ecology
  of finding resources},
 (Chapman and Hall, London 1991).

\bibitem{Oshanin} G. Oshanin, K. Lindenberg, H. S. Wio, and S. F. Burlatsky, 
 Efficient search by optimized intermittent random walks,
{\it J. Phys. A: Math. Theor.} {\bf  42}, 434008 (2009).

\bibitem{BenichouRV}  O. B\'enichou, C. Loverdo, M. Moreau, and R. Voituriez,
    Intermittent search strategies, {\it Rev. Mod. Phys.}{ \bf 83}, 81 (2011). 



\bibitem{Bartumeus} Bartumeus F and Catalan J, Optimal search behavior and classic foraging theory {\it J. Phys. A: Math. Theor.} {\bf 42} 434002 (2009)

\bibitem{Stanley}  E. R. Raposo,  S. V. Buldyrev, 
M G E dal Luz, G. M. Viswanathan, 
and H. E. Stanley,
Levy Flights and random searches
{\it J. Phys. A: Math. Theor.} {\bf 42} 434003 (2009)


\bibitem{Kafri}  Y. Kafri and R. A. da Silveira, Steady-State Chemotaxis
 in Escherichia Coli,  {\it Phys. Rev. Lett.} 
  {\bf 100}, 238101 (2008).


\bibitem{Kafri2}  M. Sheinman and  Y. Kafri, Effects of intersegmental 
 transfers of target location by  proteins,  {\it Phys. Biol.} {\bf 6},
 016003 (2009).


\bibitem{Tailleur} J. Tailleur and M. E. Cates, Statistical Mechanics of
Interacting Run-and-Tumble Bacteria, {\it Phys. Rev. Lett.} 
{\bf 100}, 218103  (2008).

\bibitem{Riggs}	A. D. Riggs, S.  Bourgeois
and M. Cohn,  The lac repressor-operator interaction. 3. Kinetic
studies, {\it J. Mol. Biol.} {\bf  53}, 401 (1970).


\bibitem{Berg} O. G. Berg, R. B. Winter and P. H. Von Hippel,
{ Diffusion-driven mechanisms of protein
translocation on nucleic acids. 1. Models and theory},
{\it Biochemistry} {\bf 20}, 6929 (1981).

\bibitem{Mirny} L. Mirny, M. Slutsky, Z.  Wunderlich, A. Tafvizi, J. Leith
  and A.  Kosmrlj,   How a protein searches for its site on DNA:
 the mechanism of facilitated diffusion,  {\it J. Phys. A: Math. Theor.}
 {\bf  42}, 434013 (2009).

\bibitem{EM1} M.~R. Evans and S.~N. Majumdar,
 Diffusion with stochastic resetting,
 {\it  Phys. Rev. Lett.}  {\bf 106}, 160601 
(2011). 

\bibitem{EM2} M.~R. Evans and S.~N. Majumdar,  Diffusion with optimal resetting,
 {\it  J. Phys. A: Math. Theor.}  {\bf 44}, 
435001 (2011).

\bibitem{MZ} S.~C. Manrubia and D.~H. Zanette, Stochastic multiplicative processes 
with reset events, {\it Phys. Rev. E} {\bf 59}, 4945 (1999).

\bibitem{MV} M. Montero and J. Villarroel, Monotonous continuous-time random walks 
with drift and stochastic reset events, Preprint arXiv: 1206.4570

\bibitem{Gelenbe} E. Gelenbe,
Search in unknown environments,
{\it Phys. Rev. E} {\bf 82}, 061112 (2010).



\bibitem{JP} S. Janson and Y. Peres,
Hitting times for random walks with restarts,
{\it SIAM J. Discrete Math.}  {\bf  26}, 537 (2012)

\bibitem{Tachiya}
M. Tachiya,
Theory of diffusion-controlled reactions:
formulation of the bulk reaction rate in
terms of the pair probability,
{\it Radiat. Phys. Chem.} {\bf 21}, 167 (1983)

\bibitem{Karlin} S. Karlin and H. E. Taylor, 
 {\it A First Course in Stochastic Processes} (Academic Press, San Diego, 1975).

\bibitem{Redner} S. Redner, {\it A guide to First-Passage Processes}
 (Cambridge University Press, Cambridge 2001).

\bibitem{PaulK}  P. L. Krapivsky, S. Redner and E. Ben-Naim,   {\it  A
Kinetic View of Statistical Physics} (Cambridge University
Press, Cambridge 2010).

\bibitem{SidPaul}  P. L. Krapivsky and  S. Redner, {Kinetics of a diffusion
 capture process:  lamb besieged by a pride of lions},
{\it  J. Phys. A: Math. Gen.}
 {\bf 29}, 5347 (1996). 

\bibitem{FM} J. Franke and S.~N. Majumdar, Survival probability of an immobile 
target surrounded by mobile traps, {\it J. Stat. Mech.} {\bf P05024} (2012).  

















\end{thebibliography}
\end{document}